\documentclass[10pt,twocolumn,letterpaper,aps,pra,longbibliography]{revtex4-1}
\usepackage[utf8]{inputenc}
\usepackage{graphicx}
\usepackage{amsmath}
\usepackage{amsfonts}
\usepackage{amsbsy}
\usepackage{eucal}
\usepackage{hyperref}

\renewcommand{\vec}[1]{\pmb{#1}}
\newcommand{\uvec}[1]{\hat{\vec{#1}}}
\newcommand{\ii}{i}
\newcommand{\ee}{e}
\newcommand{\dd}{d}
\newcommand{\pd}{\partial}
\newcommand{\bra}[1]{\langle #1|}
\newcommand{\ket}[1]{|#1\rangle}
\newcommand{\braket}[2]{{\langle#1|#2\rangle}}
\newcommand{\cg}[6]{(#1 #2 #3 #4 | #5 #6)}

\newcommand{\mc}[1]{\mathcal{#1}}
\newcommand{\sump}{\sideset{}{'}{\sum}}
\newcommand{\LSp}[3]{{}^{#1}\!{#2}^{#3}}
\newcommand{\etal}{\textsl{et al}}

\begin{document}

\title{Role of intermediate continuum states in exterior complex
scaling calculations of two-photon ionization cross section}

\author{Andrej Miheli\v{c}}
\affiliation{Jožef Stefan Institute, Jamova cesta 39, SI-1000 Ljubljana, Slovenia}

\date{\today}

\begin{abstract}

The calculation of partial two-photon ionization cross sections in the
above-threshold energy region is discussed in the framework of exterior
complex scaling.  It is shown that with a minor modification of the usual
procedure, which is based on the calculation of the outgoing partial waves
of the second-order scattering wave function, reliable partial ionization
amplitudes can be obtained. The modified procedure relies on a few-term
least-squares fit of radial functions pertaining to different partial
waves. To test the procedure, partial and total two-photon ionization
cross sections of the helium atom have been calculated for a broad range
of incident photon energies. The calculated cross sections may be seen to
agree well with the results found in the literature. Furthermore, it is shown
that, using a similar approach, partial photoionization cross sections of an
atom in an autoionizing (resonance) state may be calculated in a relatively
straightforward way. Such photoionization cross sections may find their use in
enhanced few-parameter models describing the atom-light interaction in cases
where a direct solution of the time-dependent Schr\"{o}dinger equation becomes
too resource-intensive.

\end{abstract}

\maketitle

\section{Introduction}

Continuum-continuum transitions often play an important role in
photoionization of atoms when using short-wavelength radiation from intense
coherent light sources operating in the extreme ultraviolet or x-ray spectral
regions, such as free electron lasers (FELs) or high-order harmonic generation
(HHG) sources; when the incident flux is high enough, so that the probability
for multiphoton ionization becomes non-negligible, continuum states may
be encountered as both intermediate and final states of the multiphoton
transition process. While the presence of resonance (autoionizing)
intermediate and final states poses a computational challenge, techniques
based on the method of exterior complex scaling (ECS) \cite{simon:79,
rescigno:97, kurasov:94, mccurdy:04} seem to tackle the description of both
non-structured and resonant atomic continuum in a particularly efficient and
elegant way.

ECS-based methods have been used in the calculations of ionization amplitudes
and cross sections, e.g., for one- and two-photon single and double ionization
of He \cite{mccurdy:04a, horner:07, horner:08a, horner:08b}, as well as
in time-dependent calculations, in which effective partial ionization
cross sections have been extracted from the wave packet \cite{palacios:07,
palacios:08, palacios:09}.
One of the implementations of the ECS method, the infinite-range complex
scaling (irECS) \cite{scrinzi:10}, combined with the time-dependent
surface flux approach (tSurff) \cite{tao:12}, has been used to solve
the time-dependent Schr\"{o}dinger equation on minimal simulation
volumes.  Recently, irECS has been combined with the time-dependent
complete-active-space self-consistent method \cite{sato:16} and applied to
strong-field ionization and high-harmonic generation in He, Be, and Ne atoms
\cite{orimo:18}.

The ECS method and its implementation in terms of B-splines \cite{bachau:01},
which are also used in the present work, is described in detail in Ref.\
\cite{mccurdy:04}. It is based on a transformation of radial coordinates
outside a sphere with a fixed radius ($R_0$):
\begin{equation}
  R(r) = \begin{cases}
    r &; \, r \le R_0 \\
    R_0 + (r-R_0)\ee^{\ii\theta} &; \, r > R_0
  \end{cases},
  \label{eq:trans}
\end{equation}
where $\theta > 0$ denotes the scaling angle. By applying the ECS
transformation, the Hamiltonian operator describing an atom or a molecule
becomes non-Hermitian. Requiring the wave function to vanish on the ECS
contour for $r\to\infty$, outgoing scattering boundary conditions are
imposed \cite{mccurdy:04}. Furthermore, the spectral representation of
retarded Green's operator using the eigenpairs of the transformed Hamiltonian
operator is seen to be particularly simple and convenient to implement. These
properties make the ECS method suitable for a description of the atomic and
molecular continuum and for calculations of transition (collision) amplitudes.

In this work, a procedure for the calculation of partial two-photon
ionization amplitudes and cross sections is presented. The procedure relies
on an extraction of the ionization amplitudes from the outgoing waves in the
non-scaled region of space via a least-squares fit, and is applied to the case
of two-photon ionization of He atoms. Furthermore, it is shown that a similar
procedure may be used to calculate photoionization amplitudes of an atom in a
resonance state calculated in the framework of the ECS method.

In the calculations presented in this work, 256 B-spline basis functions
\cite{bachau:01, mccurdy:04} have been used to represent the radial parts
of the single-electron wave functions. Single-electron angular momenta
up to $\ell_\mathrm{max} = 6$ have been used. Two-electron wave functions
have been written using the close-coupling approach \cite{venuti:96,
carette:13}, with the expansion augmented by either B-spline functions or
other correlation basis functions. Final-state channels with principal quantum
numbers up to $n_\mathrm{max} = 10$ have been used. Most of the calculations
have been performed using $R_0 = 80$ a.u.\ and $R_\mathrm{max} = 300$
a.u. A quadratic-linear-quadratic knot sequence has been used to achieve:
(i) an accurate description of wave functions close to the origin; (ii)
a good representation of the continuum in the non-scaled region of space;
and (iii) an adequate description of eigen wave functions of the scaled
Hamiltonian operator which are used to represent the atomic continuum for low
photoelectron kinetic energies.
Throughout this work, Hartree atomic units are used unless stated otherwise.

\section{Description of the method}

\subsection{Partial ionization amplitudes and cross sections}

Using ECS, one can calculate partial two-photon ionization amplitudes which
correspond to accessible ionization channels \cite{mccurdy:04, mccurdy:04a}.
These amplitudes are calculated from the solutions of the following set of
driven Schr\"{o}dinger equations:
\begin{align}
  &(E_0 + \omega - H) \ket{\hat\Psi_1} = D \ket{\Phi_0},
  \label{eq:def1} \\
  &(E_0 + 2\omega - H) \ket{\hat\Psi_2} = D \ket{\hat\Psi_1},
  \label{eq:def2}
\end{align}
where $H$ denotes the complex-scaled Hamiltonian operator of the free helium
atom, $\ket{\Phi_0}$ and $E_0$ the (bound) initial atomic state and its
energy, $\omega$ the photon energy, and $D$ the dipole operator. States
$\ket{\hat\Psi_1}$ and $\ket{\hat\Psi_2}$ describe the outgoing waves
of the first- and second-order scattering states usually denoted by
$\ket{\Psi^+_1}$ and $\ket{\Psi^+_2}$. A transition amplitude describing
a specific final-state channel for the case of two-photon ionization is
then calculated by analyzing the corresponding second-order wave function,
$\hat\Psi_2(\vec{r}_1,\vec{r}_2)$.

Solutions $\ket{\hat\Psi_1}$ and $\ket{\hat\Psi_2}$ are obtained by inverting
Eqs.\ \eqref{eq:def1} and \eqref{eq:def2}:
\begin{align}
  \ket{\hat\Psi_1} &= \sum_j
  \frac{\ket{\Phi_j}\bra{\Phi_j}D\ket{\Phi_0}}
       {E_0 + \omega - E_j},
       \label{eq:sol1} \\
  \ket{\hat\Psi_2} &= \sum_j
  \frac{\ket{\Phi_j}\bra{\Phi_j}D\ket{\hat\Psi_1}}
       {E_0 + 2\omega - E_j},
       \label{eq:sol2}
\end{align}
where $\ket{\Phi_j}$ and $\bra{\Phi_j}$ are the $j$th right and left
eigenvector of $H$, respectively, and $E_j$ is the (generally complex)
eigenenergy which corresponds to $\ket{\Phi_j}$. It is to be understood that
dipole matrix elements $\bra{\Phi_j}D\ket{\Phi_0}$ and
$\bra{\Phi_j}D\ket{\hat\Psi^1}$ are evaluated on the ECS contour.

For above-threshold ionization (ATI), i.e., when $E_0 + \omega$ lies in
the continuum, first-order wave function $\hat\Psi_1(\vec{r}_1, \vec{r}_2)$
describes a state in which at least one of the electrons is not bound. The
radial function associated with the continuum electron thus extends
beyond the non-scaled region of space. This makes the driving term of Eq.\
\eqref{eq:def2} $R_0$-dependent. Especially in the context of two-photon
double ionization treated in the framework of the ECS method \cite{horner:07,
horner:08a, horner:08b, morales:09}, but sometimes also for two-photon single
ionization, this may be addressed by adding a small, imaginary term $\ii\eta$
($\eta > 0$) in the denominator of Eq.\ \eqref{eq:sol1}:
\begin{align}
  \ket{\hat\Psi_1^\eta} &= \sum_j
  \frac{\ket{\Phi_j}\bra{\Phi_j}D\ket{\Phi_0}}
       {E_0 + \omega - E_j + \ii\eta},
       \label{eq:solm1} \\
  \ket{\hat\Psi_2^\eta} &= \sum_j
  \frac{\ket{\Phi_j}\bra{\Phi_j}D\ket{\hat\Psi_1^\eta}}
       {E_0 + 2\omega - E_j}.
       \label{eq:solm2}
\end{align}
The inclusion of the imaginary term results in an additional exponential
damping ($\sim \ee^{-\eta r}$) of the radial functions of the first-order
solution. By choosing a suitable value of $\eta$, the amplitudes of the radial
functions associated with continuum channels can be made negligibly small near
the boundary of the non-scaled region of space. The partial-wave amplitudes
extracted from second-order wave function $\hat\Psi^\eta_2 (\vec{r}_1,
\vec{r}_2)$ (near $r = R_0$) may be seen to vary smoothly with $\eta$ over
a relatively wide interval. This allows one to extrapolate ($\eta \to 0^+$)
their values to obtain the amplitudes of the unmodified problem.  By damping
the first-order wave function, however, the peaks which appear in the
generalized two-photon ionization cross section for $E_0 + \omega$ close to
intermediate resonance states (resonance-enhanced ionization) are artificially
broadened. Furthermore, the same applies for the contributions from the
so-called core-excited resonances \cite{shakeshaft:06}.  In these cases, the
broadening can not be ``undone'' using the limiting procedure. An alternative
way to determine the partial ionization amplitudes is discussed below.

\begin{figure}[htb!]\begin{center}
\includegraphics[width=\linewidth]{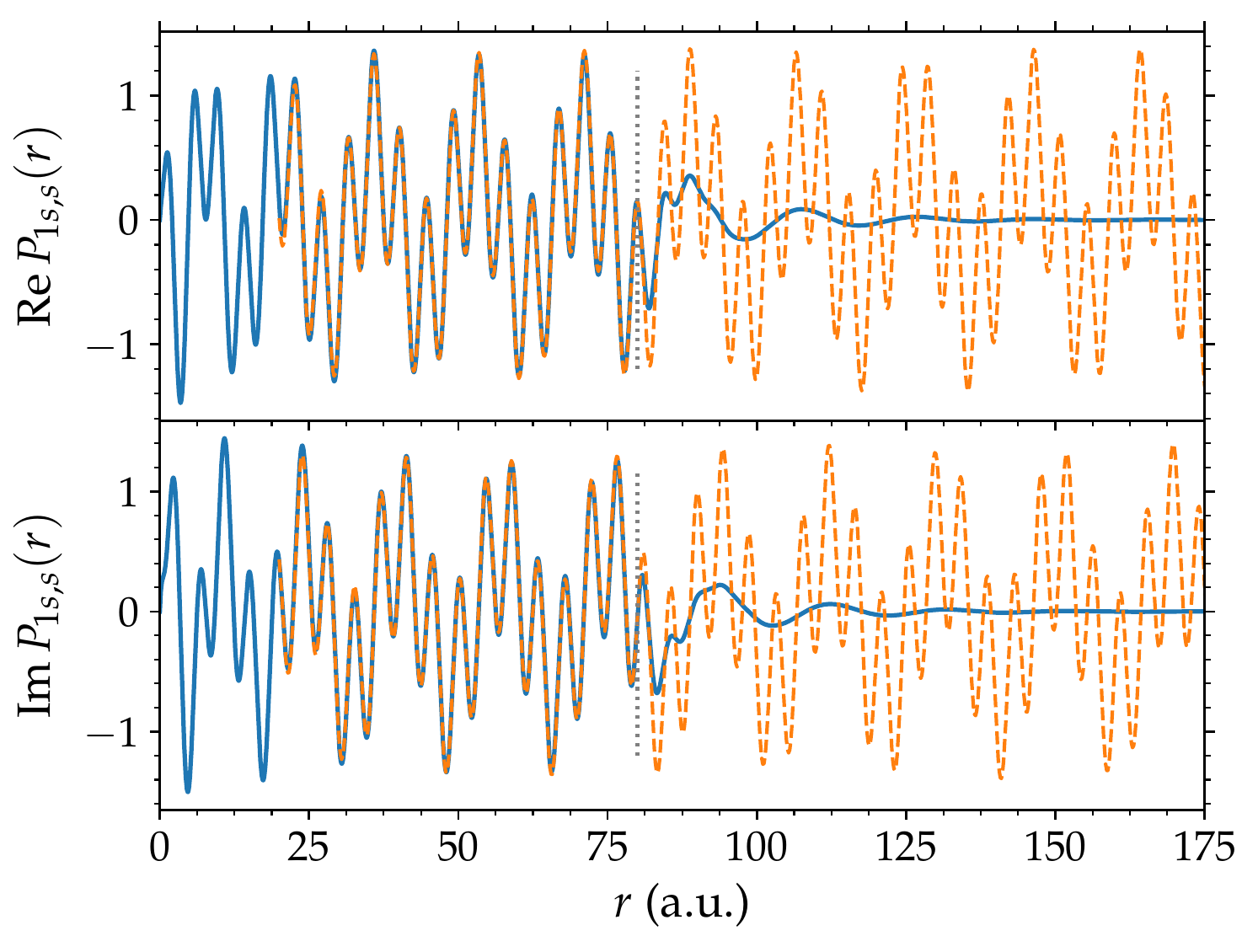}
\caption{Real (top) and imaginary (bottom) part of the radial function for
the $1s\epsilon s$ ionization channel calculated for photon energy $\omega =
0.95$ a.u.\ (solid blue lines). Parameter $R_0$ has been set to 80 a.u.\
(marked with dotted vertical lines). The result of a least-squares fit using
Eq.\ \eqref{eq:beat1} is plotted with dashed orange lines and has been
extended beyond $r = R_0$.}
\label{fig:wave1}
\end{center}\end{figure}

\begin{figure}\begin{center}
\includegraphics[width=\linewidth]{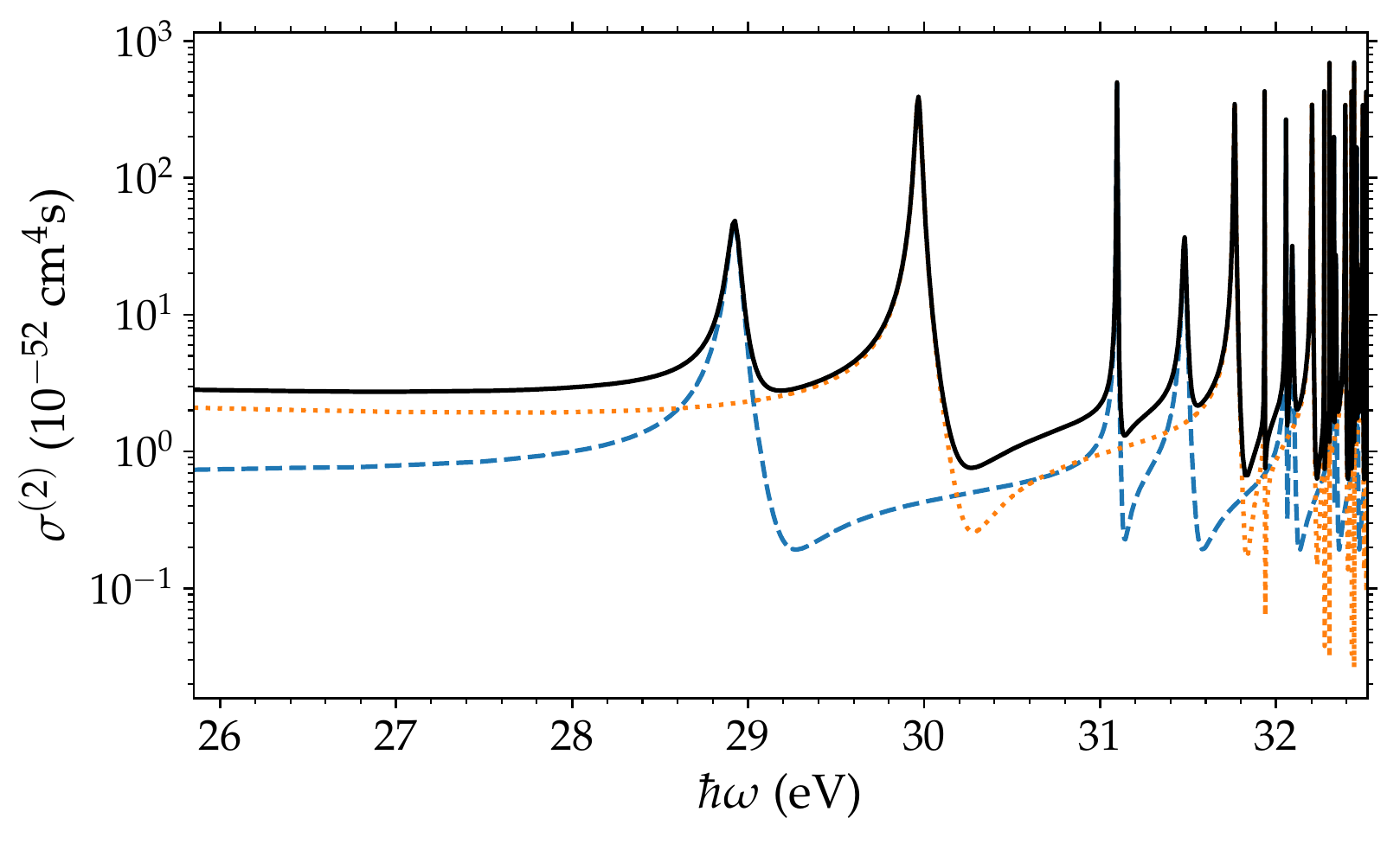}
\caption{Two-photon ionization cross section (solid black line) in the
region of the $\LSp{1}{S}{e}$ and $\LSp{1}{D}{e}$ resonance states below
the $N=2$ ionization threshold. The contributions of the $1s\epsilon s$
($\LSp{1}{S}{e}$) and $1s\epsilon d$ ($\LSp{1}{D}{e}$) channels are plotted
with a dashed blue and a dotted orange line, respectively.}
\label{fig:cs1}
\end{center}\end{figure}

Henceforth, the focus will be on the dipole operator written in the velocity
form,
\begin{equation}
  D = \uvec{e} \cdot (\vec{p}_1 + \vec{p}_2),
\end{equation}
where $\vec{p}_1$ and $\vec{p}_2$ are the electron momentum operators and
$\uvec{e}$ is the unit polarization vector. Below we show how, given this
particular form of the dipole operator, one can extract partial ionization
amplitudes from second-order state $\ket{\hat\Psi_2}$. To do this, we project
$\ket{\hat\Psi_2}$ onto a subspace spanned by the states with a fixed total
orbital angular momentum and spin, a fixed ion core, $\alpha \equiv
(n_1,\ell_1)$, and a chosen angular momentum of the remaining electron
($\ell_2$), but make no attempt to single out the partial wave with the chosen
wave number. In particular, we are not, at this stage, concerned with a
projection of $\hat\Psi_2(\vec{r}_1,\vec{r}_2)$ onto the channel wave functions
associated with specific kinetic energies of the continuum electron, which is
how partial ionization amplitudes may generally be calculated. This point will
be discussed further below. The projection, which shall be denoted by
$\ket{\hat \Psi_2^{\alpha\ell_2}}$, can be written as:
\begin{equation}\begin{aligned}
  \ket{\hat\Psi_2^{\alpha\ell_2}} &=
   \sump_\beta x_{\alpha\beta} \ket{\{\phi_\alpha\chi_\beta\}},
\end{aligned}\end{equation}
where $\ket{\{\phi_\alpha\chi_\beta\}}$ is the antisymmetric coupled
two-electron basis state with a $Z = 2$ hydrogen-like core ($\phi_\alpha$),
the primed summation runs over one-electron basis states ($\chi_\beta$)
with $\ell_\beta = \ell_2$, and $x_{\alpha\beta}$ denotes the corresponding
expansion coefficient.
Let us consider the case where the helium atom is initially in the ground state
and the photon energy is low enough, so that $E_0 + \omega$ and $E_0 + 2\omega$
fall between the first ($N=1$) and the second ($N=2$) ionization threshold.
Two-photon ionization then proceeds through the $1s\epsilon' p$ intermediate
continuum states, where the kinetic energy of the continuum electron has been
denoted by $\epsilon'$. The $1s\epsilon\ell_2$ ($\ell_2 = s, d$) continuum
channels are thus accessible in the second step, where, similarly, $\epsilon$
is the kinetic energy of the photoelectron in the final state. We write the
radial function associated with the continuum electron as:
\begin{equation}
  P_{\alpha\ell_2}(r) = \sump_\beta x_{\alpha\beta} P_{\chi_\beta}(r),
\end{equation}
where $P_{\chi_\beta}(r)/r$ is the radial part of one-electron wave function
$\chi_\beta(\vec{r})$. In the asymptotic region, $P_{\alpha\ell_2}(r)$
approximately approaches a sum of outgoing radial Coulomb functions with two
characteristic wave numbers $k$ and $k'$:
\begin{equation}\begin{aligned}
  P_{\alpha\ell_2}(r) &\sim
  \mc{B} \big\{ F_{\ell_2}(Z_c,k;r) + \ii G_{\ell_2}(Z_c,k;r) \big\} \\
  &+
  \mc{B}' \big\{ F_{\ell_2}(Z_c,k';r) + \ii G_{\ell_2}(Z_c,k';r) \big\}.
  \label{eq:beat1}
\end{aligned}\end{equation}
In Eq.\ \eqref{eq:beat1}, $F_{\ell_2}$ and $G_{\ell_2}$ are the regular and
irregular energy-normalized radial Coulomb functions, and $\mc{B}$ and
$\mc{B}'$ are the amplitudes associated with the two partial waves. In the case
of helium, $Z_c = Z - 1$, where $Z = 2$ is the nuclear charge. To see the
asymptotic form is indeed approximately given by Eq.\ \eqref{eq:beat1}, one
proceeds as follows. Firstly, the value of $k$ is fixed by the energy
conservation condition, which is a direct consequence of Fermi's golden rule:
\begin{equation}
  E_0 + 2\omega = I_{1s} + \epsilon = I_{1s} + k^2/2.
  \label{eq:erel0}
\end{equation}
Here, $I_{1s}$ is the energy of the ion core ($1s$). Secondly, the value
of $k'$ may be calculated if one takes into account that the on-shell
approximation \cite{marante:14, jimenez:16} is valid. Since this is the
case, the transition matrix elements between non-resonant (structureless)
continuum states may be seen to be approximately diagonal in the energy
\cite{jimenez:16},
\begin{equation}
  \bra{1s\epsilon' p}D\ket{1s\epsilon\ell_2}
    \sim \delta(\epsilon' - \epsilon),
  \label{eq:rel1}
\end{equation}
which leads to the following condition for the remaining wave number in Eq.\
\eqref{eq:beat1}:
\begin{equation}
  \epsilon \approx \epsilon' = E_0 + \omega - I_{1s} = k'^2/2.
  \label{eq:erel1}
\end{equation}
Amplitudes $\mc{B}$ and $\mc{B}'$ can then be extracted from
$P_{\alpha\ell_2}(r)$ by a least-squares fit, and the partial two-photon
ionization cross section of interest is seen to be proportional to
$|\mc{B}|^2$. In Fig.\ \ref{fig:wave1}, this is illustrated for the case
of the $1s\epsilon s$ ionization channel for photon energy $\omega =
0.95$ a.u.\ (25.85 eV). As can be seen, the real and imaginary parts
of $P_{\alpha\ell_2}(r) \equiv P_{1s,s}(r)$ are characterized by wave
beats. For $r \le R_0$, but at sufficiently large radii (so that the
short-range correlation potential becomes negligibly small), these beats
are accurately described by Eq.\ \eqref{eq:beat1}. It has been checked that,
although the shape of the driving term ($D\hat\Psi_1$) depends on $R_0$,
$P_{\alpha\ell_2}(r)$ remains independent of its value for $r \le R_0$. Figure
\ref{fig:cs1} shows the two-photon ionization cross section for $E_0 +
2\omega$ chosen in the region of the $\LSp{1}{S}{e}$ and $\LSp{1}{D}{e}$
autoionizing states below the $N=2$ ionization threshold. Good agreement
between the two-photon cross section calculated using the present method
and the data available in the literature has been obtained (e.g., see Ref.\
\cite{sanchez:95}).
The partial cross sections for each of the ionization channels have been
calculated using:
\begin{equation}
    \sigma^{(2)}_{\alpha\ell_2,L}(\omega) = 16\pi c^{-2}\omega^{-2}
    \sum_M \big|\mc{B}^{LM}_{\alpha\ell_2}\big|^2,
\end{equation}
where $\mc{B}^{LM}_{\alpha\ell_2}(k) \equiv \mc{B}$ is the ionization
amplitude of the channel specified by $\alpha\ell_2$, total orbital angular
momentum $L$ and its projection $M$, and wave number $k$. It has been assumed
that, asymptotically, $F_{\ell_2}(Z_c,k;r) \sim \sqrt{2/(\pi k)}
\sin\theta_{\ell_2}$ and $G_{\ell_2}(Z_c,k;r) \sim -\sqrt{2/(\pi k)}
\cos\theta_{\ell_2}$, where $\theta_{\ell_2} = kr + (Z_c/k)\ln(2kr) -
\ell_2\pi/2 + \sigma_{\ell_2}$ is the total phase and $\sigma_{\ell_2} =
\arg\Gamma(\ell_2 + 1 - \ii Z_c/k)$ is the Coulomb phase shift.

The extracted ionization amplitudes also allow one to calculate photoelectron
angular distributions (PADs). Given amplitude $\mc{B}^{LM}_{\alpha\ell_2}(k)$,
the (spin-averaged) angle-dependent ionization amplitude is given by
\cite{carette:13, okeeffe:10, okeeffe:13}:
\begin{equation}
  \mc{B}^{LM}_{\alpha\ell_2}(k) \sum_{m_2} \cg{\ell_1,}{M-m_2;}
      {\ell_2,}{m_2}{L,}{M} Y_{\ell_2m_2}(\uvec{k}).
\end{equation}
The spherical harmonic describing the angular dependence of the electron
ejection has been denoted by $Y_{\ell_2m_2}$($\uvec{k}$), and $\cg{\ell_1,}
{M-m_2;} {\ell_2,} {m_2} {L,} {M}$ is the Clebsch-Gordan coefficient for the
coupling between the angular momentum of the ion core and the angular momentum
of the continuum electron. The calculation of the PADs and the corresponding
asymmetry parameters for one- and two-photon ionization is described in detail
in Refs.\ \cite{carette:13, okeeffe:10, okeeffe:13}.

As the photon energy is further increased, but is low enough so that $E_0 +
\omega$ still lies between the $N=1$ and $N=2$ thresholds, additional
final-state continuum channels become accessible, e.g., $2s\epsilon s$,
$2p\epsilon p$ ($\LSp{1}{S}{e}$ and $\LSp{1}{D}{e}$), etc.  Let us look at the
calculation of the $2p\epsilon p$ partial ionization amplitudes. The equality
for $k$ now reads:
\begin{equation}
  E_0 + 2\omega - I_{2p} = k^2/2.
  \label{eq:erel0x}
\end{equation}
In this case, however, the relation between $\epsilon$ and $\epsilon'$
analogous to Eq.\ \eqref{eq:rel1} is seen to be a consequence of the property
of the dipole matrix element for the $1s\epsilon' p \to 2p\epsilon p$
continuum-continuum transition, in which the continuum electron acts as a
spectator:
\begin{equation}
  \bra{1s\epsilon' p}D\ket{2p\epsilon p} \propto
  \braket{\epsilon' p}{\epsilon p} \propto \delta(\epsilon' - \epsilon).
\end{equation}
This leads to the condition
\begin{equation}
  \epsilon = \epsilon' = k'^2/2.
  \label{eq:erel2}
\end{equation}
As before, the relevant amplitude ($\mc{B}$) is determined by a least-squares
fit.

Finally, when the photon energy is even further increased, several (say,
$K$), channels are open at energy $E_0 + \omega$ (the first step), and
$P_{\alpha\ell_2}(r)$ is written as a sum of at most $K+1$ terms of the form
given in Eq.\ \eqref{eq:beat1}. The number of terms depends on the number
of allowed continuum-continuum transitions ($n'_1\ell'_1\epsilon'\ell'_2 \to
n_1\ell_1\epsilon\ell_2$). The fitting procedure has been found to be stable,
as long as $K$ has remained reasonably small. For photon energies above the
second ionization threshold, the degeneracy in the intermediate step (e.g., for
the $2s\epsilon' p$, $2p\epsilon' s$, and $2p\epsilon' d$ ionization channels)
has been handled by solving the normal equations using a pseudo-inverse.

\begin{figure}\begin{center}
\includegraphics[width=\linewidth]{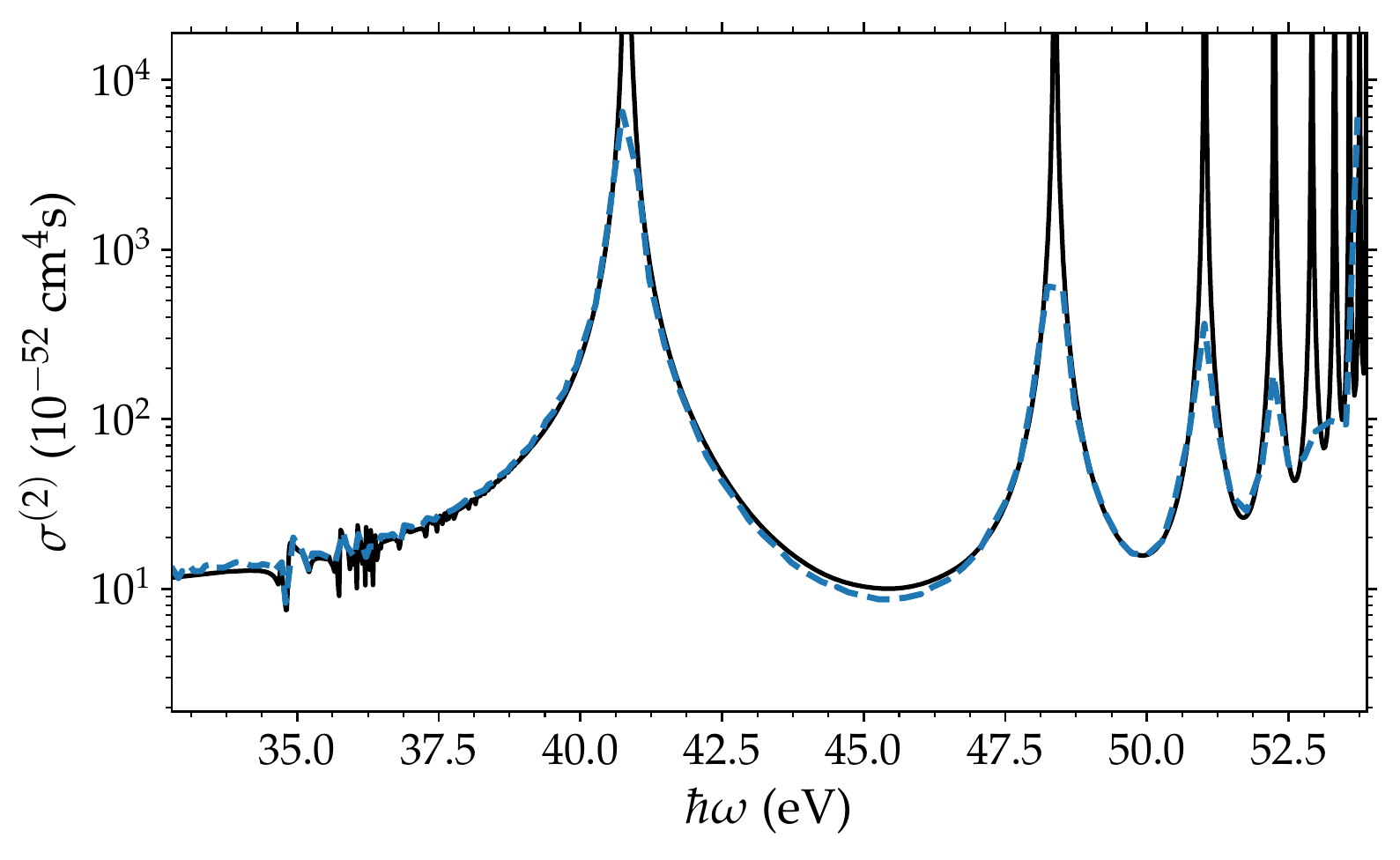}
\caption{Two-photon ionization cross section in the region of the
core-excited resonances. The result obtained using a least-squares fit with
$R_0 = 80$ a.u.\ (solid black) and the result of Shakeshaft
\cite{shakeshaft:06} (dashed blue) are shown.}
\label{fig:cs2}
\end{center}\end{figure}

It is interesting to analyze the behavior of $P_{\alpha\ell_2}(r)$ when $k$
coincides with one or several other wave numbers in the sum. This situation may
occur, for example, for two-photon ground-state ionization through the
$1s\epsilon' p$ states in the case of the $2p\epsilon p$ channels discussed
above. The resulting wave numbers are equal when
\begin{equation}
  \epsilon = E_0 + \omega - I_{1s} = E_0 + 2\omega - I_{2p} = \epsilon',
  \label{eq:erel3}
\end{equation}
which holds for $\omega = I_{2p} - I_{1s}$. When the photon energy lies close
to $I_{2p} - I_{1s}$, the normal equations of the least-squares problem become
ill-conditioned. (We discuss this further below.) This results in a resonant
enhancement in the two-photon partial cross section, which is a signature of
the core-excited resonance \cite{shakeshaft:06}. A similar behavior is also
present at higher photon energies, specifically, when the photon energy
equals $I_{n_1p} - I_{1s}$. It is important to note at this point that the
core-excited resonances are accessible via continuum-continuum transitions
in a neutral atom (i.e., not an ion). In the present case, the relevant
transitions are of the form
\begin{equation}
  1s\epsilon' p \to n_1 p\epsilon p,
\end{equation}
for which the continuum electron does not actively participate, as has already
been mentioned. Furthermore, as has been argued by Shakeshaft
\cite{shakeshaft:06}, the present formalism for the description of two-photon
ionization, in which the field-dressing (broadening) effects have not been
taken into account, is not adequate for photon energies which lie very close to
the positions of the core-excited resonances. A similar behavior is encountered
below the ionization thresholds, when $E_0 + \omega$ coincides with the
energies of the bound $1snp$ states, unless their decay widths are taken into
account.
Figure \ref{fig:cs2} shows the total two-photon ionization cross section
for photon energies above the $N=2$ threshold. Apart from the contributions
of the final resonance states converging to higher ionization thresholds
(for $\hbar\omega$ around 35 eV), a series of spikes due to the core-excited
resonances is visible (centered at approximately 40.8 eV, 48.4 eV, 51.0 eV,
etc.). The present result is seen to agree well with the result of Ref.\
\cite{shakeshaft:06}.

The present approach allows one to accurately treat correlation in the initial,
intermediate, and final states. In particular, electron correlation in the
first- and second-order solutions, $\ket{\hat\Psi_1}$ and $\ket{\hat\Psi_2}$,
is taken into account through correlated eigenstates $\bra{\Phi_j}$ and
$\ket{\Phi_j}$ in Eqs.\ \eqref{eq:sol1} and \eqref{eq:sol2}. In the present
case, electron correlation has been taken into account by including correlation
basis states in the close-coupling expansion. In the current implementation,
given an expansion which contains correlation wave functions and $n_1\ell_1
\epsilon\ell_2$ continuum channel wave functions with energies of the ion core
$I_{n_1\ell_1}$ up to $E_c$, partial ionization amplitudes for photon energies
$\omega \le (E_c - E_0)/2$ can be calculated. Conversely, when the final-state
energy lies above the threshold for double electron ejection, all the
single-ionization channels are open. A pure close-coupling expansion has been
used in this case.

\begin{figure}\begin{center}
\includegraphics[width=\linewidth]{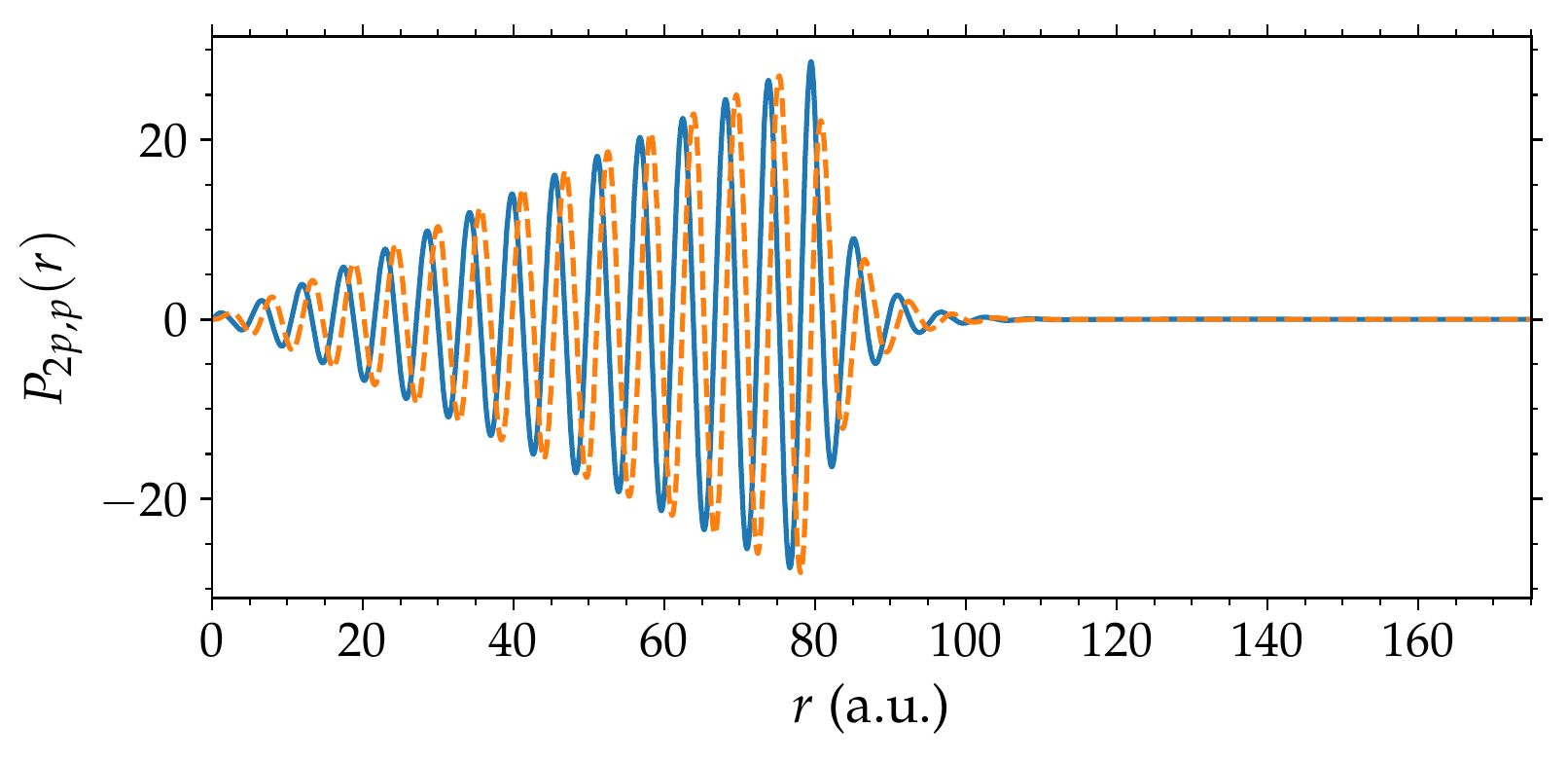}
\caption{Radial function $P_{2p,p}(r)$ for the case of $\omega = I_{2p} -
I_{1s} = 0.5$ a.u. The real and imaginary part are plotted with a solid blue
and dashed orange line, respectively.}
\label{fig:wave2}
\end{center}\end{figure}

Another comment is in place here concerning the structure of Eq.\
\eqref{eq:def2}. When the photon energy lies above the ionization threshold,
the radial part associated with the driving term in Eq.\ \eqref{eq:def2} has a
``harmonic-like'' form (the form of an outgoing Coulomb wave). The radial part
of the Schr\"{o}dinger equation thus, loosely speaking, resembles the equation
of motion of a forced harmonic oscillator, with the independent variable
replaced with radius $r$. For $\omega = I_{2p} - I_{1s}$ and without any
additional damping, wave number $k'$ (the ``driving frequency'') matches $k$
(the ``frequency of the oscillator''). This case corresponds to a resonantly
driven (non-damped) harmonic oscillator. An analogous behavior for the radial
function pertaining to the $2p\epsilon p$ $\LSp{1}{S}{e}$ channel can be seen
in Fig.\ \ref{fig:wave2}: the amplitude of the radial function increases
monotonically for $r \le R_0$.

\begin{figure}\begin{center}
\includegraphics[width=\linewidth]{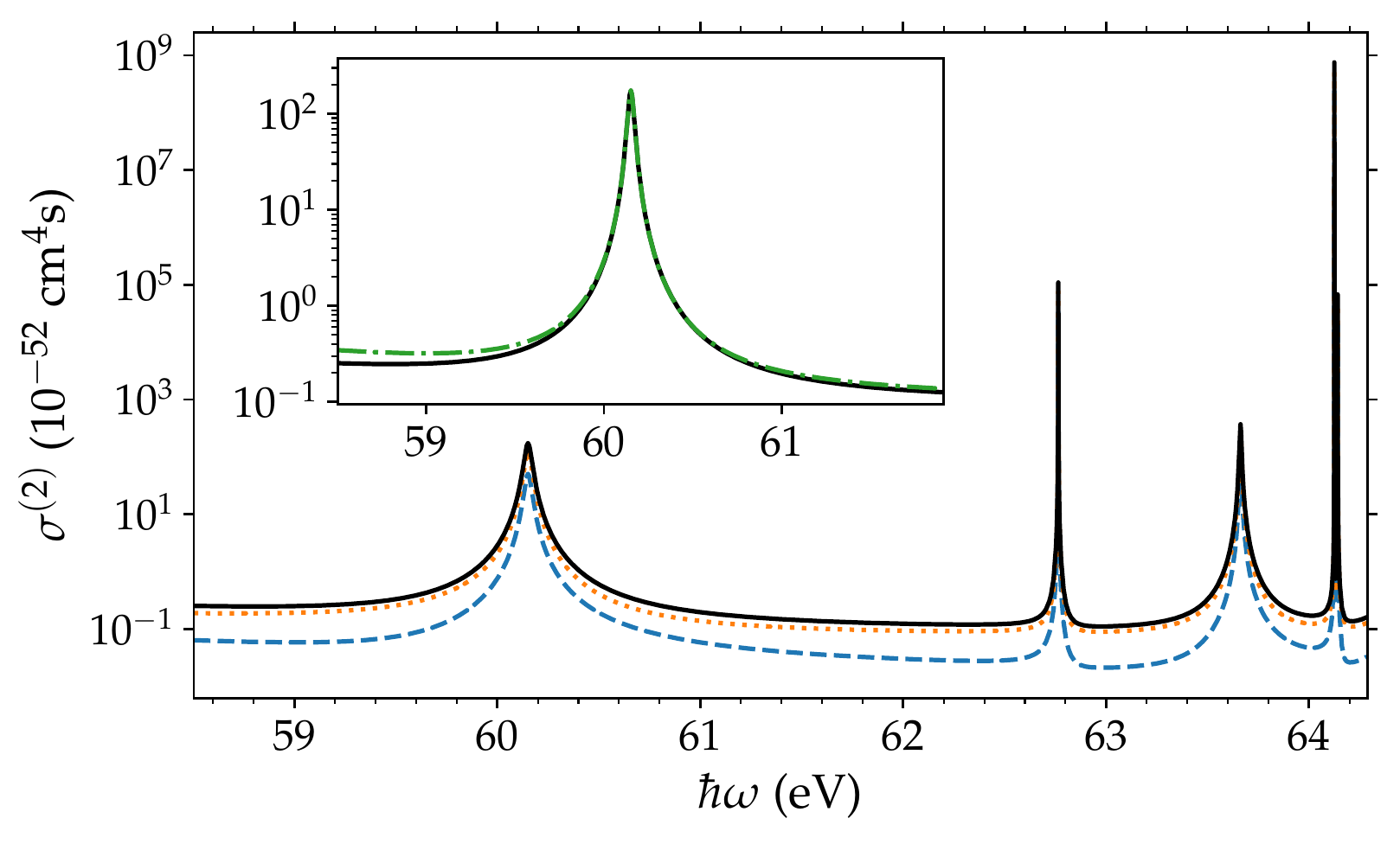}
\caption{Resonance-enhanced two-photon ionization cross section in the
energy region of the lowest $\LSp{1}{P}{o}$ autoionizing states. The dashed
blue and dotted orange line show the contributions of the $\LSp{1}{S}{e}$
and $\LSp{1}{D}{e}$ final-state ionization channels. The total cross section
is plotted with a solid black line. In the inset, the total cross section in
the region of the $sp_2^+$ $\LSp{1}{P}{o}$ doubly-excited state is compared to
the result obtained using the projection onto the final-state channel
functions (green dash-dotted line), as described in Section
\ref{sec:projection}.}
\label{fig:csrempi}
\end{center}\end{figure}

As has been mentioned, the two-photon ionization cross section is enhanced
when $E_0 + \omega$ lies close to the energies of the intermediate resonance
states, which is usually referred to as resonance-enhanced ionization. The
resonantly enhanced two-photon ionization cross section is depicted in
Fig.\ \ref{fig:csrempi} for the case of the lowest ($N=2$) $\LSp{1}{P}{o}$
intermediate autoionizing states \cite{cooper:63, rost:97}. Since these
states lie between the first and the second ionization threshold, only
the $1s\epsilon p$ continuum is open at $E_0 + \omega$. As before,
the wave numbers associated with the final-state ionization channels
$n_1\ell_1\epsilon\ell_2$ are determined from $k^2/2 = E_0 + 2\omega -
I_{n_1\ell_1}$, whereas for the intermediate step, $k'^2/2 = E_0 + \omega -
I_{1s}$ is used. Similar results have also been obtained for higher-lying
resonance states, including those with the energies converging to the $N
= 3$ ionization threshold. In the inset of Fig.\ \ref{fig:csrempi}, the
total cross section in the region of the $sp_2^+$ ($2^+$) $\LSp{1}{P}{o}$
state \cite{cooper:63} is compared to the cross section calculated
using the projection to the channel wave functions (see Section
\ref{sec:projection}). The minor differences between the results calculated
with the two approaches are due to the change in the background caused by the
tails of the broadened nearby peaks of the core-excited resonances.

The least-squares procedure is also applicable in the case of two-color
driving. In order to calculate the corresponding two-color ionization
amplitudes, $\omega$ and $2\omega$ in Eqs.\ \eqref{eq:sol1}, \eqref{eq:sol2},
\eqref{eq:erel0}, \eqref{eq:erel1}, \eqref{eq:erel0x}, and \eqref{eq:erel3}
should be replaced with photon energies of the two sources, $\omega_1$ and
$\omega_2$. Depending on these photon energies, the resulting two-photon
ionization amplitudes may be seen to correspond to the different cases studied
in Ref.\ \cite{jimenez:16}: (i) the case of resonance-enhanced photoionization,
where the final-state continuum is non-resonant; (ii) the case of two-photon
ionization which proceeds through a non-resonant continuum and where the energy of
the final state lies in the region of a resonance state; and (iii) the general
case of doubly-resonant ionization.

\subsection{Projection onto the final-state channel functions}
\label{sec:projection}

It has been mentioned that the partial ionization amplitudes can be calculated
by projecting second-order wave function $\hat\Psi_2(\vec{r}_1, \vec{r}_2)$
onto the channel functions which describe the continuum electron with
specific kinetic energy $\epsilon$.  When the ECS method is used, the
projection integral is limited to the non-scaled region of space ($r \le
R_0$). Equivalently, the integral can be transformed to a surface integral if
the form of the channel (``testing'') functions is chosen appropriately (e.g.,
see Refs.\ \cite{mccurdy:04a, palacios:07, horner:08a}). Also in this case,
the non-scaled spatial region is involved. While the peaks due to the
intermediate bound or resonance states which appear in the two-photon cross
section are not affected by the integration over the finite volume, the latter
results in a broadening of the peaks due to the core-excited resonances. This
broadening may be determined from the smallest difference between $k$ and $k'$
which can still be resolved using the projection integral. Since $0\le r \le
R_0$, this difference is given by $\Delta k = |k-k'| \sim 2\pi/R_0$. For
$\Delta k$ small compared to $k$ and $k'$, the broadening may be assessed
from:
\begin{equation}
  \Delta \epsilon \approx \sqrt{2\epsilon} \, \Delta k
  \sim \sqrt{2\epsilon} \, \frac{2\pi}{R_0}.
  \label{eq:broad}
\end{equation}
Conversely, when the partial amplitudes are extracted using a fit, no
additional broadening occurs. The advantage of the least-squares fit procedure
is thus that it allows one to extract the partial ionization amplitudes even
when $R_0$ is relatively low, as long as the shape of the radial function close
to $r = R_0$ is adequately described by Eq.\ \eqref{eq:beat1} or its
multi-term generalization.

\begin{figure}\begin{center}
\includegraphics[width=\linewidth]{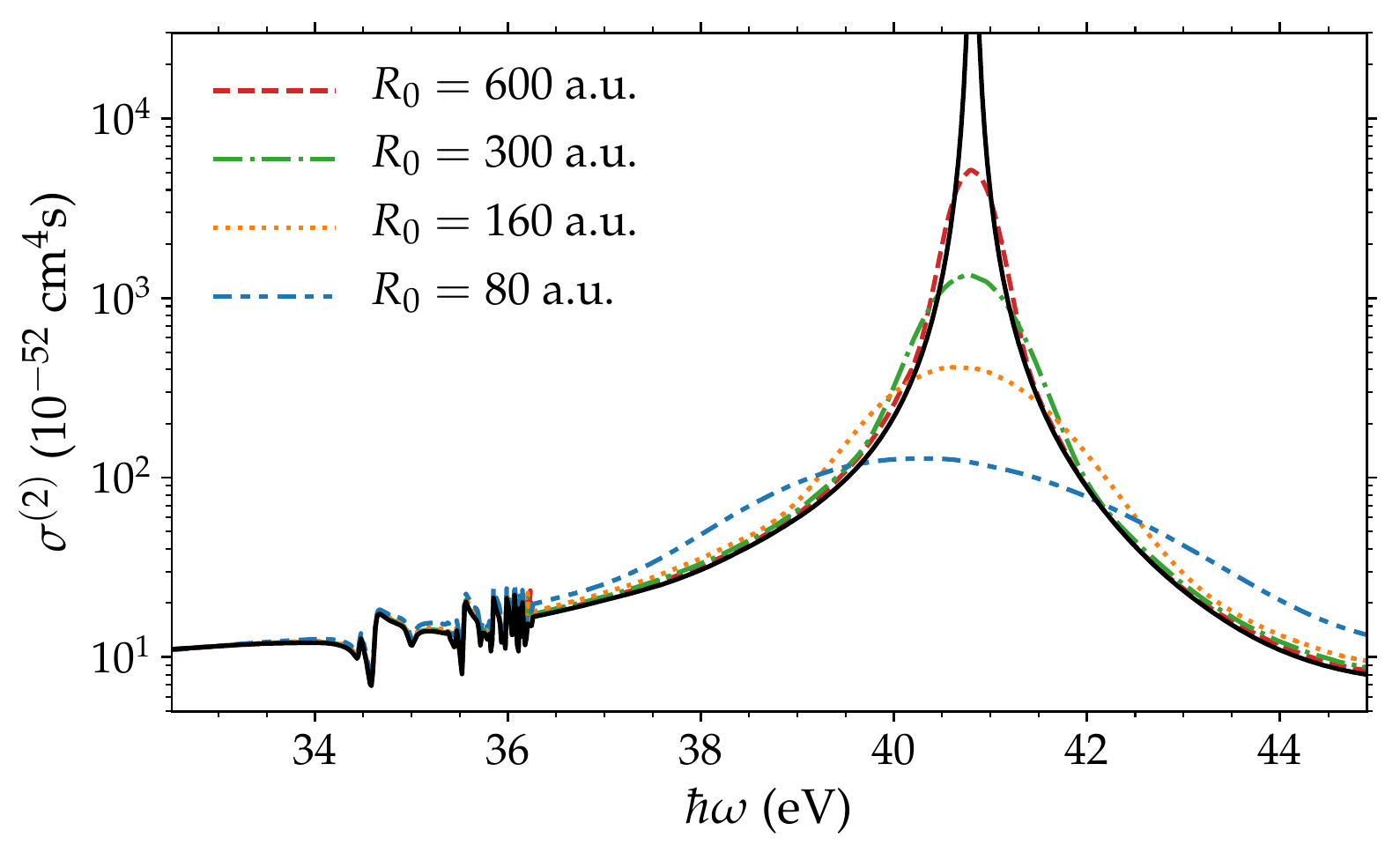}
\caption{Two-photon ionization cross section for different scaling radii
$R_0$ calculated with the projection approach. The cross section calculated with the least-squares fit has been
plotted with a solid black line.}
\label{fig:csint}
\end{center}\end{figure}

The partial two-photon ionization cross sections have been calculated by means
of a projection for comparison. The ionization amplitude of the $\alpha\ell_2$
channel has been calculated using the following approach:
\begin{align}
   \mc{B}_{\alpha\ell_2} &= \frac{\mc{I}[P_{\alpha\ell_2}(r)]}
       {\mc{I}[H_{\ell_2}(Z_c,k;r)]},
   \label{eq:proj1} \\
   \mc{I}[ f(r) ] &= \int_{R_\mathrm{min}}^{R_0}
   \big\{H_{\ell_2}(Z_c,k;r)\big\}^\ast f(r) \, w(r) \dd r,
   \label{eq:proj2}
\end{align}
where $H_{\ell_2}(Z_c,k;r) = F_{\ell_2}(Z_c,k;r) + \ii G_{\ell_2}(Z_c,k;r)$,
and a window function, denoted by $w(r)$ in Eq.\ \eqref{eq:proj2}, has been
added under the integral to reduce the oscillatory artifacts. These appear
due to the finite-interval (Fourier-like) side-lobes which accompany the peaks
when the integrand is non-zero at the upper integral bound.
The total two-photon ionization cross section in the energy region of the
lowest core-excited resonance ($I_{2p} - I_{1s} = 0.5 \text{ a.u.} \approx
40.8$ eV) calculated with the least-squares fit (for $R_0 = 80$ a.u.) and the
projection (for $R_0 = 80$ a.u., 160 a.u., 300 a.u., and 600 a.u.)
is shown in Fig.\ \ref{fig:csint}. These results have been obtained for
$R_\mathrm{min} = 5$ a.u.\ and by setting $w(r) = \exp\{-(\kappa r)^n\}$,
with $n = 4$. For each of the values of $R_0$, parameter $\kappa$
has been calculated by requiring that $w(R_0) = 0.001$. It should be noted
here that the effect of the window function is fundamentally different from
the effect of the damping term ($\ii\eta$) in Eq.\ \eqref{eq:solm1}. While
in the former case, the widths of the peaks due to the intermediate bound and
resonance states remain unchanged, the peaks are artificially broadened using
the latter approach.
As expected, the agreement between the cross section calculated with the
least-squares fit and the cross sections calculated with the projection
approach is better for higher values of the $R_0$ parameter. It has been
found that in some cases, like for the $2s\epsilon s$ channel, the partial
wave amplitude approaches the asymptotic value slowly with increasing $r$ if
the photon energy lies close to the energy of the core-excited resonance. In
these cases, larger $R_0$ values have to be considered even when the amplitude
is extracted using the fit.

Equations \eqref{eq:proj1} and \eqref{eq:proj2} result in two-photon
ionization cross sections which are almost indistinguishable from those
calculated with the least-squares fit for photon energies chosen in the energy
regions of intermediate bound and resonance states (as shown in the inset of
Fig.\ \ref{fig:csrempi}).

\begin{figure}\begin{center}
\includegraphics[width=\linewidth]{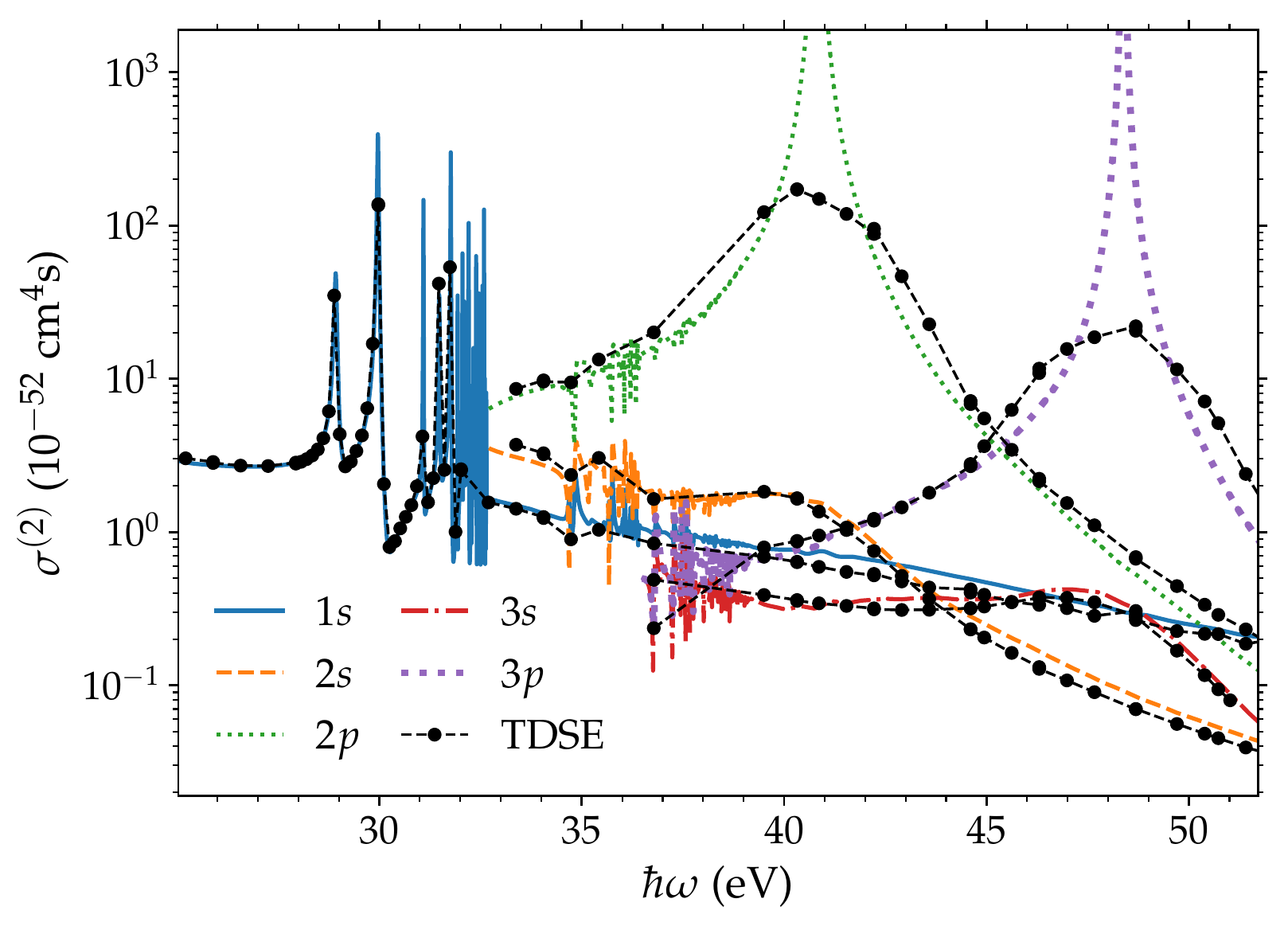}
\caption{Partial two-photon ionization cross sections as given by Eq.\
\eqref{eq:csion}. The $1s$, $2s$, $2p$, $3s$, and $3p$ partial cross sections
are shown. The results of the time-dependent calculation (TDSE) for 2 fs long
pulses (from Ref.\ \cite{palacios:09}) are plotted with black circles.}
\label{fig:csion}
\end{center}\end{figure}

A similar procedure based on a projection onto a set of channel functions is
also used to extract partial single- and double-ionization amplitudes from a
solution of the time-dependent Schr\"{o}dinger equation (TDSE). The extraction
of the ionization amplitudes from a time-dependent wave packet using the ECS
method has been treated extensively by Palacios \etal.\ \cite{palacios:07,
palacios:08, palacios:09}. Let $\ket{\Psi(t)}$ denote the solution of the TDSE,
\begin{equation}
    \big\{H + V(t)\big\}\ket{\Psi(t)} = \ii \frac{\pd}{\pd t}\ket{\Psi(t)},
\end{equation}
where $V(t)$ describes the interaction of the atom with an electromagnetic
pulse of duration $\tau$. At the end of the pulse, the outgoing part of the
wave function may be calculated from
\begin{equation}
    (E - H)\ket{\hat\Psi} = \ket{\Psi(\tau)},
\end{equation}
which has the form of Eqs.\ \eqref{eq:def1} and \eqref{eq:def2}. Partial
ionization amplitudes can then be extracted from wave function $\hat
\Psi(\vec{r}_1, \vec{r}_2)$, which allows one to calculate the corresponding
partial ionization cross sections. As can be seen in Fig.\ \ref{fig:csion},
there is good overall agreement between the partial two-photon ionization cross
sections calculated with the present method and the cross sections from Ref.\
\cite{palacios:09}, which have been obtained by solving the TDSE for 2 fs long
pulses. Each of the cross sections in Fig.\ \ref{fig:csion} describes an
ionization process which leads to the helium ion in a specific state. These
cross sections have been calculated by summing up the partial cross sections
pertaining to ionization channels with a fixed ion core ($\alpha$), but with
different angular momenta of the continuum electron ($\ell_2$) and different
values of the total angular momentum ($L = 0$ or $L = 2$ for linearly polarized
light in the present case):
\begin{equation}
    \sigma^{(2)}_\alpha(\omega)
        = \sum_{\ell_2,L} \sigma^{(2)}_{\alpha\ell_2,L}(\omega).
    \label{eq:csion}
\end{equation}
The sharp peaks due to the core-excited resonances are broader in the TDSE
case. As has already been mentioned, the field-broadening effects have not been
taken into account in the present calculations. Furthermore, in the
time-dependent treatment, additional broadening occurs due to the finite
excitation bandwidth (i.e., due to the finite pulse duration). Specifically,
the widths of the peaks in the TDSE case may be seen to be inversely
proportional to the radial extent of the resulting wave packet (i.e., not to
$R_0$, as in the present case). Note, however, that in Ref.\
\cite{palacios:09}, the value of $R_0$ has been chosen high enough, so that the
outgoing wave packet can be assumed not to have reached the boundaries of the
non-scaled spatial region for time $t \le \tau$. Since this is the case, Eq.\
\eqref{eq:broad} may still be used to give the lower bound for the spectral
broadening.

Let us conclude by noting that the extraction of partial ionization amplitudes
from the solution of the TDSE, i.e., from $\Psi(\vec{r}_1,\vec{r}_2;\tau)$, is
generally not possible using the least-squares fit procedure described in this
work. The reason for this is that radial function $P_{\alpha\ell_2}(r)$ can not
generally be written in the form of Eq.\ \eqref{eq:beat1} or its
generalization, i.e., with a sum over a discrete set of wave numbers
($k,k',\ldots$). Instead, the general form for $P_{\alpha\ell_2}(r)$ in the
asymptotic region may be seen to be:
\begin{equation}
  \int \dd{}k \, \mc{B}_{\alpha\ell_2}(k)
  \big\{ F_{\ell_2}(Z_c,k;r) + \ii G_{\ell_2}(Z_c,k;r) \big\}.
  \label{eq:Pgen}
\end{equation}
The projection approach described above may in this case be used to extract
$\mc{B}_{\alpha\ell_2}(k)$. The same limitations of course apply for the
calculation of the PADs and energy spectra of the ejected electrons in the
time-dependent framework.
To this end, the following is to be noted. Energy-resolved partial two-photon
ionization cross sections may be trivially calculated in the framework of the
time-independent perturbation theory, and are seen to be proportional to the
modulus square of the relevant partial amplitudes obtained with the
least-squares fit:
\begin{equation}
  \frac{\dd \sigma^{(2)}_{\alpha\ell_2,L}}
    {\dd E} \propto \sum_M \big|\mc{B}^{LM}_{\alpha\ell_2}\big|^2
    \, \delta(E_0 + 2\omega - E).
  \label{eq:sige}
\end{equation}
In Eq.\ \eqref{eq:sige}, $E = I_\alpha + \epsilon$ is the energy of the final
state. This is in contrast to the time-dependent treatment, for which the
partial amplitude can not generally be extracted from the wave packet using a
fit, and photoelectron energy spectra may only be calculated by projecting
$\Psi(\vec{r}_1,\vec{r}_2;\tau)$ onto the channel functions describing the
continuum electron with a fixed kinetic energy ($\epsilon$).

\subsection{Gauge invariance}

\begin{figure}\begin{center}
\includegraphics[width=\linewidth]{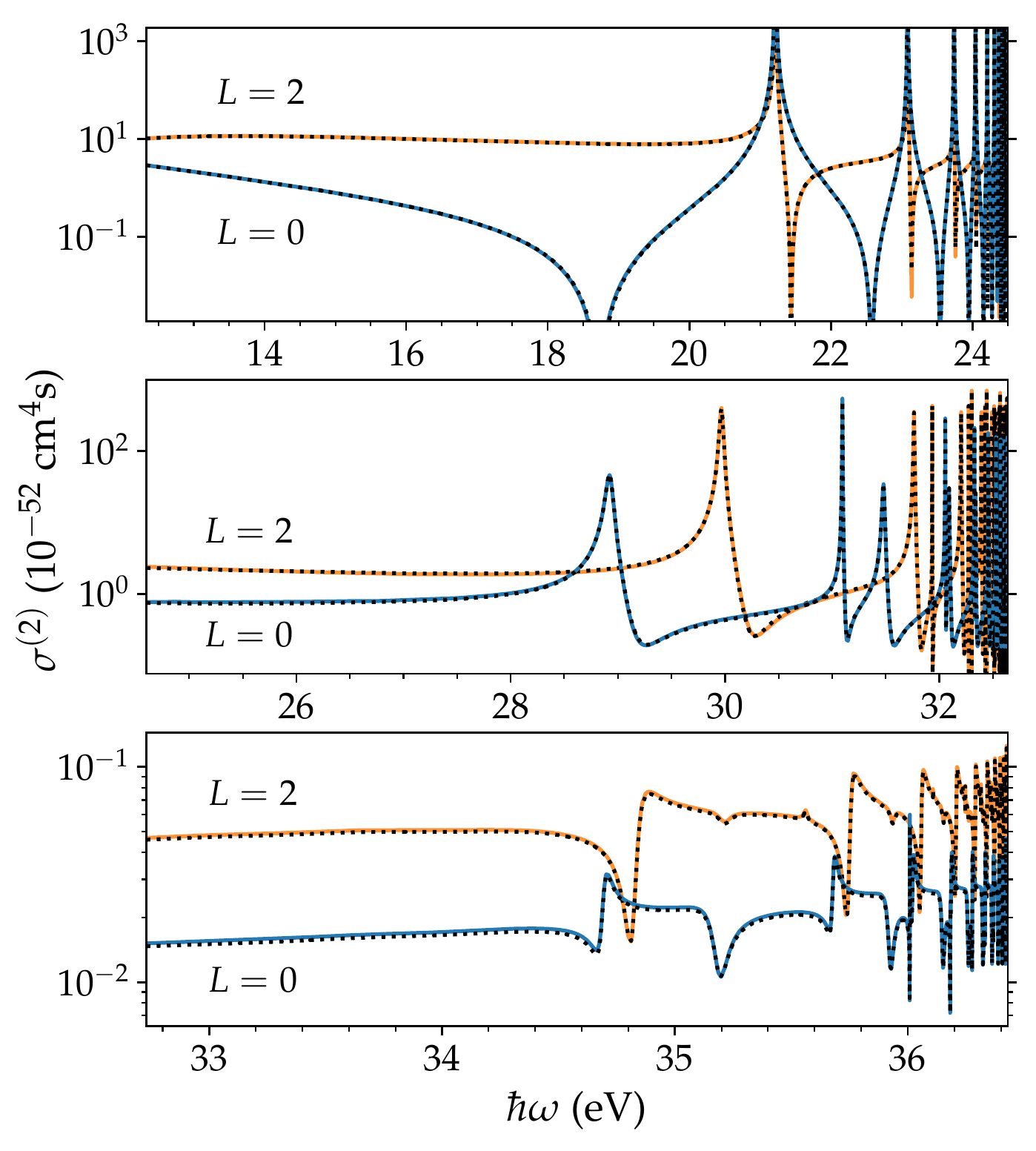}
\caption{The $\LSp{1}{S}{e}$ ($L = 0$) and $\LSp{1}{D}{e}$ ($L = 2$) two-photon
ionization cross section below the $N=1$ (top), $N = 2$ (middle), and $N = 3$
(bottom) ionization thresholds. The results for the length- and velocity-form
dipole operator are plotted with solid and dotted lines, respectively.}
\label{fig:csLV}
\end{center}\end{figure}

Perhaps surprisingly, good agreement has been obtained between the cross
sections calculated using the velocity-form and those employing the length form
of the dipole operator, $D = \uvec{e} \cdot (\vec{r}_1 + \vec{r}_2)$, where
$\vec{r}_1$ and $\vec{r}_2$ denote the position operators of the two electrons.
The dipole matrix elements between the eigenstates of $H$ have been transformed
to the velocity form using the well-known relation
\begin{equation}
   \bra{\Phi_a}\vec{p}_1 + \vec{p}_2\ket{\Phi_b} =
   \ii (E_a - E_b) \bra{\Phi_a}\vec{r}_1 + \vec{r}_2\ket{\Phi_b}
   \label{eq:relLV}
\end{equation}
prior to calculating the first- and second-order solutions
[Eqs.\ \eqref{eq:sol1} and \eqref{eq:sol2}]. In Eq.\ \eqref{eq:relLV},
eigenenergies $E_a$ and $E_b$ correspond to eigenstates $\ket{\Phi_a}$ and
$\ket{\Phi_b}$, respectively. Note that Eq.\ \eqref{eq:relLV} holds for exact
eigenstates, so a discrepancy between the two forms serves as a measure of
accuracy of their numerical representations. Arguably, the above relation only
holds for the off-the-energy-shell matrix elements \cite{marante:14}. Note,
however, that the eigenstates with different total angular momentum and parity
which are used to represent the atomic continuum are non-degenerate in the
present calculations.  The non-degeneracy is connected to the finite radial
interval used to represent the radial functions; the eigenvalues pertaining to
the ``box-normalized'' states differ. The two-photon cross sections calculated
using the length- and velocity-form of the dipole operator are shown in
Fig.\ \ref{fig:csLV}.

An analogous transformation of the transition matrix elements from the
acceleration to the velocity form results in spurious oscillations in the
two-photon cross section even for photon energies below the first ionization threshold. The oscillations are most
probably due to modifications of the commutation relations \cite{marante:14,
mercouris:94} connecting the acceleration and velocity forms of the dipole
operator, which would need to be taken into account when dealing with
box-normalized states, but have not been included in the present tests.

\section{Photoionization of an atom in a resonance state}

Theoretical treatment of photoexcitation and photoionization with
short-wavelength radiation by a direct solution of the time-dependent
Schr\"{o}dinger equation presently becomes prohibitively lengthy as soon as
the pulse duration exceeds a couple of tens of femtoseconds. When this is the
case, the time-dependent description of atom-photon interaction is usually
limited to a restricted subset of basis states by means of which the main
features of the system can be described. This includes finding the solution
of the time-dependent Schr\"{o}dinger equation in terms of the time-dependent
amplitudes of the basis states from the restricted space \cite{madsen:00,
themelis:04, mihelic:17}, studying the dynamics in terms of the density
matrix (e.g., see Refs.\ \cite{zitnik:14, mihelic:15}), or solving a set of
kinetic equations which describe the population of various atomic and ionic
species during the interaction with the incident pulse \cite{makris:09,
ilchen:16}. The parameters which enter the model, such as autoionization
widths, asymmetry parameters, photoionization cross sections, and Rabi
frequencies, can be conveniently calculated using the ECS method. In this
section, we show how to meaningfully define and calculate the photoionization
cross section of an atom in a resonance state.

In the framework of the ECS method, resonance (autoionizing) states are
associated with the discrete part of the eigen-spectrum of complex-scaled (ECS)
Hamiltonian operator $H$, i.e., with the complex poles of the resolvent, $G(z)
= (z-H)^{-1}$. Let $\ket{\Phi_0}$ denote an eigenstate of $H$, which represents
a resonance state. Its energy can be shown to be $\theta$-independent and may
be written as $E_0 = E_0^r - \ii\Gamma_0/2$, where $E_0^r$ and $\Gamma_0$
denote the energy position of the resonance and its decay (autoionization)
width, respectively. When the resonance is narrow, i.e., when its
autoionization width is small, so that $E_0$ lies close to the real axis, the
resonance state may be treated as a non-decaying state. In this case, we may
replace $E_0$ in the denominator of Eq.\ \eqref{eq:sol1} with $E_0^r$:
\begin{equation}
  \ket{\hat\Psi_1} = \sum_j
  \frac{\ket{\Phi_j}\bra{\Phi_j}D\ket{\Phi_0}}
       {E_0^r + \omega - E_j}.
       \label{eq:solx1}
\end{equation}
As before, states $\bra{\Phi_j}$ and $\ket{\Phi_j}$ denote the left and right
eigenvectors of $H$, $E_j$ is the eigenenergy corresponding to $\ket{\Phi_j}$,
and matrix element $\bra{\Phi_j}D\ket{\Phi_0}$ is evaluated on the ECS
contour.

\begin{figure}\begin{center}
\includegraphics[width=\linewidth]{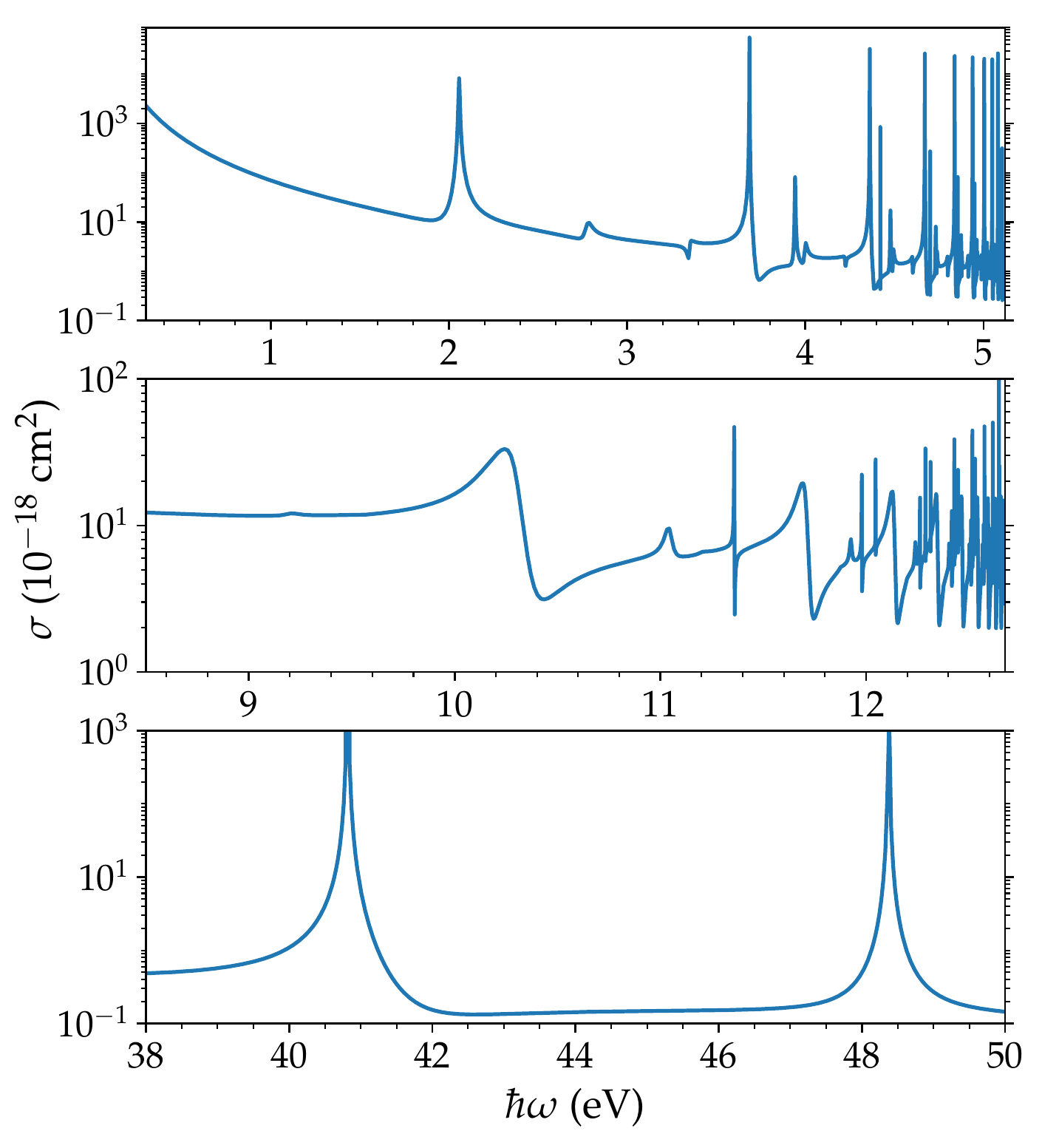}
\caption{The photoionization cross section of the He atom in the $sp_2^+$
$\LSp{1}{P}{o}$ autoionizing state for the incident photon energy below the
$N=2$ ionization threshold (top), between the $N=2$ and $N=3$ thresholds
(middle), and in the region of the core-excited resonances (bottom).}
\label{fig:csres}
\end{center}\end{figure}

Let us start by assuming that $\ket{\Phi_0}$ describes the $sp_2^+$
$\LSp{1}{P}{o}$ autoionizing state, which lies below the $N = 2$ ionization
threshold. In a similar way as before, wave number $k$ is determined from:
\begin{equation}
  E_0^r + \omega = I_{n_1\ell_1} + \epsilon = I_{n_1\ell_1} + k^2/2,
  \label{eq:relx1}
\end{equation}
where, again, $\alpha = (n_1,\ell_1)$ are the quantum numbers of the ion core
for a chosen final-state channel (described by $\alpha\ell_2$). Contrary to the
bound initial state, the $sp_2^+$ resonance state lies above the $N = 1$
threshold, and thus the $1s\epsilon' p$ continuum is open at energy $E^r_0$.
Since the resonance state also contains a small admixture of the continuum
\cite{rost:97}, the relation analogous to Eq.\ \eqref{eq:erel1} now reads
\begin{equation}
  \epsilon \approx \epsilon' = E_0^r - I_{1s} = k'^2/2.
  \label{eq:relx2}
\end{equation}
Exactly as before, radial function $P_{\alpha\ell_2}(r)$ may be calculated
from first-order state $\ket{\hat\Psi^{\alpha\ell_2}_1}$, which, in turn, is
obtained by projecting the solution of Eq.\ \eqref{eq:solx1} to the relevant
subspace. In the asymptotic region, the behavior of $P_{\alpha\ell_2}(r)$ is
approximately described by Eq.\ \eqref{eq:beat1}. In this way, the partial
ionization amplitude for the $\alpha\ell_2$ channel can be extracted. This
procedure can be generalized to higher-lying resonance states.
To demonstrate its applicability, we have calculated the partial
photoionization cross sections for an atom initially in the $sp_2^+$
$\LSp{1}{P}{o}$ autoionizing state. The results are shown in Fig.\
\ref{fig:csres} for three energy regions: below the second ionization
threshold, between the second and third threshold, and close to the
lowest-lying core-excited resonances. As can be seen, the cross section in the
latter region is enhanced in a similar way as in the case of two-photon
ionization due to the continuum-continuum transitions. This enhancement occurs
due to the admixture of the $1s\epsilon'p$ continuum, i.e., due to the
$1s\epsilon'p \to np\epsilon p$ transitions. As has been the case for the
two-photon ionization cross section, the field-dressing effects have not been
taken into account.

\section{Conclusion}

A slightly modified procedure for the calculation of partial two-photon
ionization amplitudes and cross sections based on the method of exterior
complex scaling (ECS) has been presented. The procedure relies on an
extraction of the amplitudes from radial functions of outgoing scattered waves
obtained by fixing the state of the ion core and the angular momentum of the
continuum electron.  The amplitudes are not calculated by projecting out the
partial waves associated with a fixed value of the kinetic energy of the electron;
instead, the extraction is implemented by means of a few-term linear
least-squares fit. As has become customary in the framework of the ECS method,
the scattered wave is calculated by solving a set of driven Schr\"{o}dinger
equations. While for photon energies above the ionization threshold, the first-order
driving term depends on the scaling radius, the second-order scattered wave
has been seen to be independent of the scaling parameters in the non-scaled
region of space. In a sense, this region contains a ``complete'' information
on the photoionization process, and because of this, the least-squares
fit allows for a relatively straightforward extraction of the photoionization amplitudes in the case of the
single-electron ejection.  Finally, basing on similar theoretical grounds,
a method for the calculation of partial photoionization amplitudes for an atom
in an autoionizing state has been proposed, which may be useful when a direct
solution of the Schr\"{o}dinger equation is unfeasible, and one resorts to
modeling using a restricted set of states.

\begin{acknowledgments}
The author acknowledges the financial support from the Slovenian Research
Agency (research program No.\ P1--0112). This work was supported by the
European COST Action XLIC CM1204.
\end{acknowledgments}

\bibliographystyle{apsrev4-1}
\bibliography{main}

\end{document}